# Effect of Signal Quantization on Performance Measures of a 1st Order One-Dimensional Differential Microphone Array

Shweta Pal*
*Centre for Applied Research in Electronics,*
*Indian Institute of Technology Delhi*
New Delhi, India
Shweta.Pal@care.iitd.ac.in
*Corresponding author

Arun Kumar
*Centre for Applied Research in Electronics,*
*Indian Institute of Technology Delhi*
New Delhi, India
arunkm@care.iitd.ac.in

Monika Agrawal
*Centre for Applied Research in Electronics,*
*Indian Institute of Technology Delhi*
New Delhi, India
maggarwal@care.iitd.ac.in

*Abstract*—In practical systems, recorded analog signals must be digitized for processing, introducing quantization as a critical aspect of data acquisition. While prior studies have examined quantization effects in various signal processing contexts, its impact on differential microphone arrays (DMAs), particularly in one-dimensional (1D) first-order configurations, remains unexplored. This paper investigates the influence of signal quantization on the performance of first-order 1D-DMAs across various beampatterns. An analytical expression for quantized beamformed output for a first-order 1D-DMA has been formulated. The effect of signal quantization has been studied on array performance measures such as the Beampattern, Directivity Factor (DF), Front-to-Back Ratio (FBR), and suppression depth at null points (SDN). Simulation results reveal that the beampattern shape remains structurally invariant across quantization bit depths, with quantization primarily affecting SDN. DF and FBR remain constant with the varying number of quantization bits. Additionally, SDN is shown to be frequency-independent, however, it increases with increasing quantization bit depths, enhancing interference suppression. The study also examines the effect of steering nulls across the azimuthal range, showing that SDN degrades as the null moves closer to the source's look direction ($0°$), indicating reduced interference suppression.

I. INTRODUCTION

In a wide range of speech-centric applications, such as teleconferencing systems, smart voice-controlled assistants, hearing aids, in-car voice communication, and automatic speech recognition (ASR) interfaces, microphone arrays play a crucial role in enhancing the intelligibility and quality of the desired speech signal. These arrays are deployed to spatially filter acoustic signals and extract the target speech source from complex acoustic environments typically characterized by ambient noise, competing interference sources, and reverberation. The core signal processing technique enabling this functionality is beamforming [1]. It involves the spatially selective processing of recorded multichannel signals across multiple microphone elements to enhance the desired source signal while attenuating unwanted components such as noise and interference [2-5]. Among the various beamforming techniques developed over the past few decades [6-9], differential beamforming, which forms the basis of differential microphone arrays (DMAs), has gained significant attention, especially for applications requiring compact and portable configurations [10]. DMAs are special arrays that are designed to capture the spatial derivatives of the sound pressure field, with the finite difference between adjacent microphones' outputs effectively approximating these pressure gradients. These differential arrays are particularly advantageous as they can provide frequency-invariant beampattern and achieve high directional gain using smaller array apertures [11-16].

In most practical scenarios, signals are initially recorded in analog form, but digital systems such as computers, DSPs, and microcontrollers require these signals to be represented in digital format for further processing. This necessitates analog-to-digital conversion, where quantization plays a key role by mapping continuous signal amplitudes to discrete levels. The authors in [17]-[18] analyse the impact of quantization on a binaural hearing aid device using various beamforming schemes in terms of different performance measures. In another study [19], the authors investigated the impact of phase quantization on the performance of analog beamforming in large-scale aperture array radio telescopes. They proposed a methodology to evaluate beam pointing accuracy and sidelobe levels, providing insights for a cost-effective design through both simulation and prototype measurements.

Although prior research has extensively studied the impact of signal quantization on conventional beamforming and array processing systems, the specific effects of quantization on differential array architectures, such as DMAs, have not been well investigated, especially in terms of performance and spatial filtering characteristics. This work aims to systematically analyse the role of signal quantization in a first-order one-dimensional (1D) DMA by formulating a mathematical expression for the quantized beamformed output. Specifically, the study investigates how signal quantization affects key array performance measures, including beampattern, directivity factor (DF), and front-to-back ratio (FBR), across various 1st order fixed beampatterns of dipole, cardioid, hypercardioid, and supercardioid. Here, DF measures the array's ability to focus its radiated energy in a particular direction relative to an isotropic radiator, while FBR quantifies the ratio of signal strength in the desired (front) direction to that in the opposite (back) direction [14]. The simulation results demonstrate that the beampattern shape across different fixed patterns remains structurally consistent with quantization. However, quantization predominantly affects the suppression depth at null points (SDN). SDN is the amount of signal attenuation (in dB) at a null direction in the beampattern, indicating how effectively the array suppresses signals arriving from that angle. It quantifies the amount of suppression offered to interfering signals coming from null directions. Ideally, a null in the array beampattern corresponds to a point where the array response drops to zero, resulting in SDN $=-\infty$ dB. However, in practice, SDN values are finite and can deviate significantly from this ideal due to factors such as DMA pattern type, DMA order and the quantization bit resolution. SDN values for different 1st order 1D-DMA patterns are presented as a function of varying quantization bits. Increasing the number of quantization bits increases the value of SDN. Furthermore, it was shown that both DF and FBR were independent of the changing quantization bit depths. The results also indicate that SDN remains invariant across frequency, demonstrating the frequency-robust nature of null suppression in the 1D-DMA configuration. However, a notable improvement in SDN is observed with increasing quantization bit depths. The study also examines the effect of steering the null across the azimuthal range ($0°$ to $180°$). Results show that SDN degrades as the null approaches the source look direction (i.e. $0°$), indicating reduced interference suppression.

The remainder of this paper is structured as follows. In Section II, we describe the signal model. Section III formulate the mathematical expression for the quantized beamformed signal output for a first-order 1D-DMA. Simulation results are presented in Section IV, followed by the relevant discussion of the findings. Finally, conclusions are drawn in Section V.

## II. SIGNAL MODEL

We consider a narrowband source signal $s(t)$ propagating through a free-field and an acoustic environment at the speed of sound $c$. The source signal is made incident on a 1D-DMA setup as shown in Fig. 1, where $\delta$ is the inter-element spacing considered between two adjacent sensors. The direction of the incoming source signal arrival is denoted by $\theta_a$, with $\theta_a = 0°$ representing the end-fire direction. Sensor-1 is considered the reference sensor.

The source signal incident on the reference sensor is given by:

$$s(t) = B \cos(\omega_0 t + \varphi_{sig}) \quad (1)$$

where $B$, $\varphi_{sig}$ and $\omega_0$ denotes amplitude, initial phase, and angular frequency respectively, of the incoming source signal.

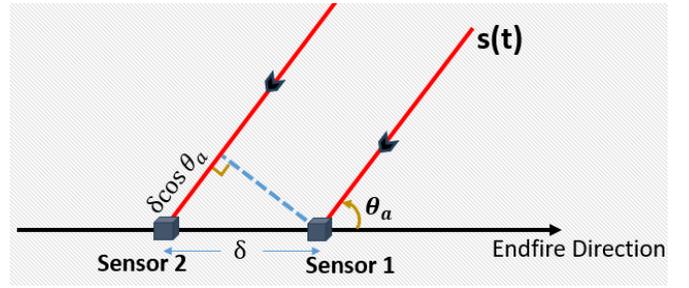

Fig. 1. Signal Model for a first order 1D-DMA.

Considering the phase and gain mismatch between adjacent sensors, the transfer function $T$ in polar form, relating sensor gain $G_s$ and phase $\phi_s$ at $\omega_0$ is given by:

$$T(\omega_0) = G_s(\omega_0) e^{j\phi_s(\omega_0)} \quad (2)$$

For a 1st order 1D-DMA, the time domain signal received at sensor-2 is then given by:

$$r(t) = s(t - \zeta_0).T(\omega_0) \quad (3)$$

where $\zeta_0 = \delta \frac{\cos(\theta_a)}{c}$ is the time delay between adjacent sensors at the arrival angle $\theta_a$.

## III. QUANTIZATION ANALYSIS

The time domain analog signal is sampled and quantized by a data acquisition (DAQ) system to produce a discrete time, digitally represented signal. Thus, the digitally sampled and quantized version of the signal, sampled at frequency $f_s$ is represented as:

$$r^q[n] = Q\{G_s(\omega_0) B \cos(\omega_0(n - N_0)T_s + \varphi_{sig} + \phi_s(\omega_0))\}. \quad (4)$$

where $N_0 = \frac{\zeta_0}{T_s}$ is the fractional delay corresponding to $\zeta_0$ in the discrete domain and $T_s = \frac{1}{f_s}$ is the sampling time. The function $Q\{.\}$ represents the quantized form of its input. Here $r^q[n]$ is the in-phase signal component equivalently denoted by $r^q_{in}[n]$. Let $r^q_{quad}[n]$ denote the corresponding quadrature phase signal of $r^q[n]$ defined as:

$$r^q_{quad}[n] = Q\{G_s(\omega_0) B \sin(\omega_0(n - N_0)T_s + \varphi_{sig} + \phi_s(\omega_0))\}. \quad (5)$$

The digital signal can be expressed in complex form using its in-phase and quadrature-phase signal components as follows:

$$z^q[n] = r^q_{in}[n] + j r^q_{quad}[n]. \quad (6)$$

Complex weights $H$ are applied to the outputs of each recorded sensor signal, and the weighted signals are then summed together to produce beamformed output of DMA. Each complex weight $H$ is product of two components: the first represents component due to the complex beamforming filter weights, while the second accounts for the complex weight due to known inverse transfer function at the angular frequency $\omega_0$ to compensate for the gain and phase mismatch between the sensors. Thus,

$$H = (W\,e^{j\Psi})\left(\frac{1}{G_s(\omega_0)}e^{-j\phi_s(\omega_0)}\right)$$
$$= \frac{W}{G_s(\omega_0)}\{\cos(\Psi - \phi_s(\omega_0)) + j\sin(\Psi - \phi_s(\omega_0))\}. \quad (7)$$

In this setup, only the sensor signal is quantized into an integer or fixed-point format, as typically done by a DAQ system. The filter coefficients and all arithmetic operations, however, are performed using the processor or software's native data type, such as float or double precision.

The complex output signal, obtained after applying the weights, is thus expressed as:

$$U^q[n] = H\,z^q[n]. \quad (8)$$

The quantized beamformed output can then be expressed as:

$$b_o^q[n] = Re\{U^q[n]\} = \left[\frac{W}{G_s(\omega_0)}\cos(\Psi - \phi_s(\omega_0))\right]\left[Q\{G_s(\omega_0)\,B\cos(\omega_0(n - N_0)T_s + \varphi_{sig} + \phi_s(\omega_0))\}\right] - \left[\frac{D}{G_s(\omega_0)}\sin(\Psi - \phi_s(\omega_0))\right]\left[Q\{G_s(\omega_0)\,B\sin(\omega_0(n - N_0)T_s + \varphi_{sig} + \phi_s(\omega_0))\}\right] \quad (9)$$

where $Re\{.\}$ represents the real part of the input argument.

The quantized signal can be represented as the sum of the original (unquantized) signal and the associated quantization error. Consequently, the beamformed output derived from the quantized signal can be expressed as the sum of two components: (a) terms resulting from the processing of the unquantized signal, and (b) terms arising due to the propagation of quantization error through the processing chain. The expression enclosed in [.] indicates the contributions from the unquantized component.

Thus, (9) can be rewritten for the quantized beamformed output of a first-order 1D-DMA as:

$$b_o^q[n] = \begin{bmatrix} B\,W\cos(\omega_0(n - N_0)T_s + \varphi_{sig} + \Psi) + \\ \frac{W}{G_s(\omega_0)}\cos(\Psi - \phi_s(\omega_0))\,\mathcal{E}_{in}[n] - \\ \frac{D}{G_s(\omega_0)}\sin(\Psi - \phi_s(\omega_0))\,\mathcal{E}_{quad}[n] \end{bmatrix} \quad (10)$$

where $\mathcal{E}_{in}[n]$ and $\mathcal{E}_{quad}[n]$ are the quantization errors associated with the in-phase signal component and the quadrature phase signal component. The quadrature-phase component may be derived from the in-phase component in the analog domain or in the digital domain. Both $\mathcal{E}_{in}[n]$ and $\mathcal{E}_{quad}[n]$ are considered independent and are reasonably well modelled using uniform probability density functions having pdf's as $U\left(-\frac{\Delta_{in}}{2}, \frac{\Delta_{in}}{2}\right]$ and $U\left(-\frac{\Delta_{quad}}{2}, \frac{\Delta_{quad}}{2}\right]$ respectively. Here $\Delta_{in}$ and $\Delta_{quad}$ represents the quantization step-size for the in-phase and quadrature phase signal components respectively. The quadrature phase signal component of received signal $r^q_{quad}[n]$ can be generated via analog circuitry prior to digitization or through Hilbert transform digitally implemented in the processor.

The expression for the beampattern ($BP$) is then given as:

$$BP(\theta_a) = \frac{1}{P}\sum_n |b_o^q[n]|^2 \quad (11)$$

where $n = 1, 2, \ldots\ldots..P$ where $P$ is the length of sequence.

IV. SIMULATION RESULTS AND DISCUSSIONS

In this section, we present simulation results for signal quantization study for a first-order 1D-DMA and study the performance with three widely used performance measures, viz. beampattern, DF and FBR. SDN variation with frequency, quantization bit depths, and null steering has also been studied. Simulations have been carried out at source frequency = 1999 Hz. We choose a sensor spacing between adjacent sensors to be $\delta = 0.04\,\lambda = 6.9$cm (assuming $c = 343$ m/s). The sampling frequency considered in the simulations is 44.1kHz. In each Monte Carlo run, $\varphi_{sig}$ and $\phi_s$ are randomly generated between 0 to $2\pi$. A total of 5000 Monte Carlo simulations were conducted to compute the results across various 1st order patterns.

Fig. 2 shows polar plot representation for beampatterns corresponding to various fixed configurations of a first-order 1D-DMA, namely dipole, cardioid, hypercardioid, and supercardioid, each characterized by a single null steered at $\theta_{dip} = 90°$, $\theta_{card} = 180°$, $\theta_{hyp} = 120°$, $\theta_{sup} = 135°$ respectively. The plots are generated using 16-bit quantized input signals. The simulation results show that quantization of the signal does not affect the overall shape of the pattern, i.e. pattern shape across different fixed patterns remain structurally consistent with quantization, thereby preserving the directional characteristics. However, quantization predominantly affects the SDN for different patterns. Table 1 provides a summary of the SDN values (in dB) corresponding to various patterns of a first-order 1D-DMA, evaluated at a quantization depth of 16 bits.

TABLE I. SDN values for 1st order 1D-DMA with 16-bit quantization

| S. No. | Beampattern | SDN (in dB) |
|---|---|---|
| 1. | Dipole | -83.1 |
| 2. | Cardioid | -88.5 |
| 3. | Hypercardioid | -85.1 |
| 4. | Supercardioid | -86.2 |

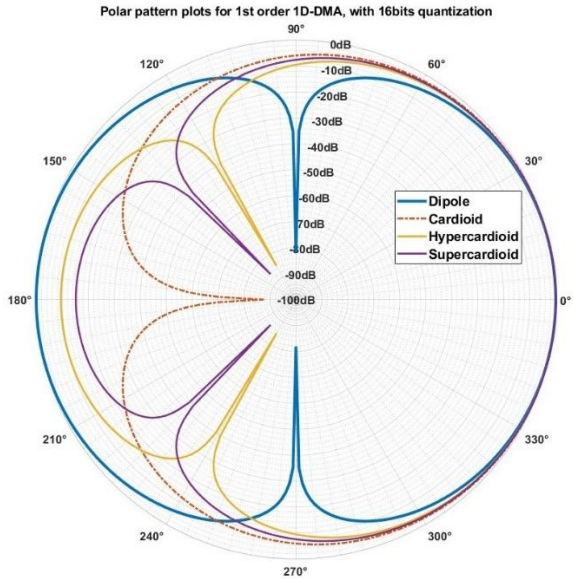

Fig. 2. Polarplot representation for $1^{st}$ order 1D-DMA.

Fig. 3. and Fig. 4. show the variation of DF and FBR, respectively, with the number of quantization bits for different $1^{st}$ order DMA beampatterns viz dipole, cardioid, hypercardioid, and supercardioid. While the frequency invariance of DF and FBR has been established in previous studies [10], the present work extends this analysis by demonstrating their invariance with respect to the number of quantization bits. This behaviour is theoretically consistent, as quantization primarily impacts the SDN rather than altering the mainlobe characteristics or the overall shape of the beampattern. Given that both DF and FBR are performance measures that predominantly depend on the mainlobe properties, they remain unaffected by changes in quantization bit depths.

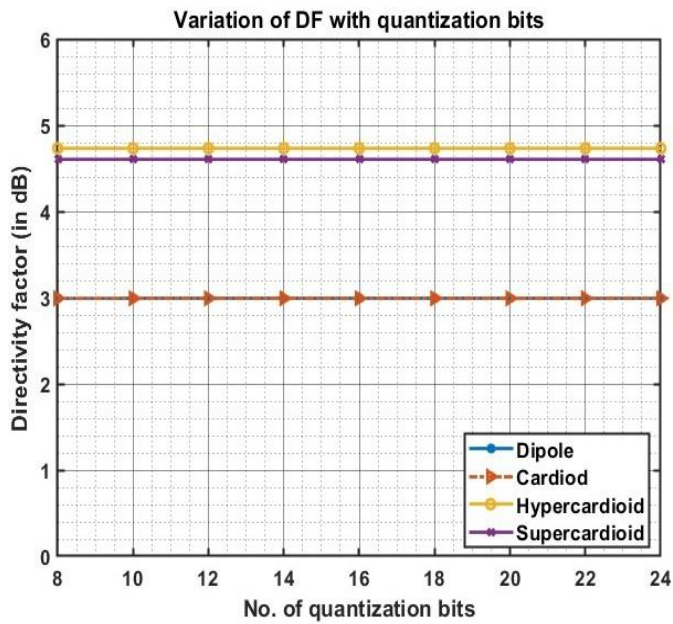

Fig. 3. DF variation with quantization bits for 1D-DMA

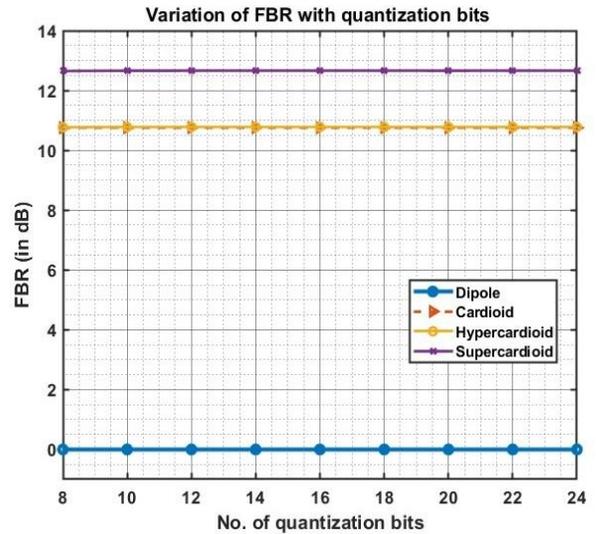

Fig. 4. FBR variation with quantization bits for 1D-DMA

Fig. 5. shows the variation of SDN with respect to both frequency as well as quantization bit depth for a $1^{st}$ order 1D-DMA dipole pattern case. The analysis spans a frequency range from 1 kHz to 6 kHz, and quantization bits variation is shown for 10, 12, 14, and 16 bits respectively. The results show that SDN remains invariant across the frequency range, indicating consistent null suppression performance, with only a minor deviation of about 2 dB observed at higher frequencies around 4 kHz for 16-bit quantization depth. As the number of quantization bits increases, SDN improves significantly, from approximately -46.9 dB at 10 bits to as deep as -83.1 dB at 16 bits. The depth values are defined w.r.t maxima of the beampattern taken as 0 dB.

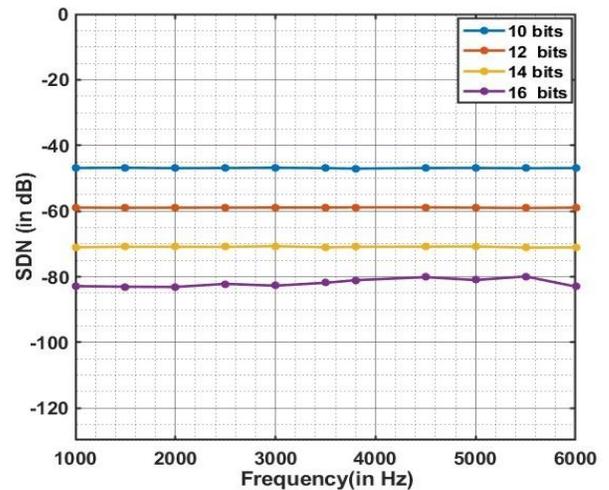

Fig. 5. Variation of SDN with frequency and quantization bits for 1D-DMA dipole pattern

Fig. 6. illustrates the variation in SDN as the null is steered across the azimuthal range. (Considering the spatial symmetry of the beampattern about the $0°$-$180°$ axis, the result is shown from $0°$ to $180°$). It is observed that SDN degrades as the null is steered closer to the source look direction ($0°$). The maximum SDN or the highest suppression is achieved when the null is positioned at 180°, corresponding to a cardioid response, where the interfering source is most angularly separated from the desired source. As the null angle approaches 0°, the suppression depth continues to degrade. SDN varies significantly from -88.5 dB (at 180°) to -46.3 dB (at 1°), as the null is steered from 180° toward 0°. Given that the desired source is positioned at 0° (endfire direction), this analysis highlights how the relative position of the null to the source direction significantly influences suppression performance. Results in Fig. 6 are shown for a quantization depth of 16 bits.

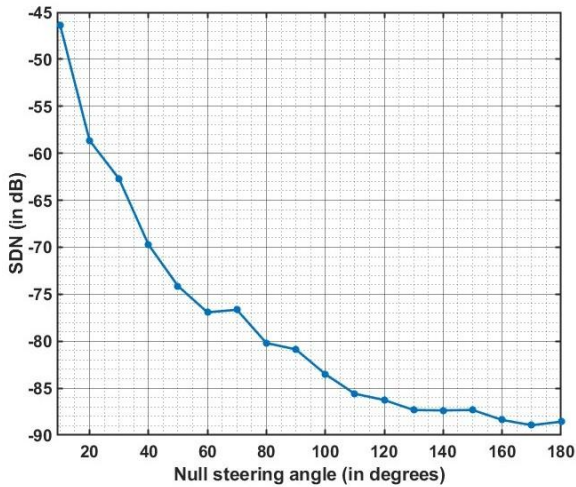

Fig. 6. Variation of SDN with null steering angle for 1D-DMA.

## V. CONCLUSION

The process of quantization that is essential in digitizing analog signals, has been widely studied, but its impact on DMAs remains largely unexplored. This paper examines how signal quantization affects various DMA beampatterns and presents an analytical expression for first-order 1D-DMA quantized beamformed output. We validated the performance analysis of the 1D-DMA through simulations using the beampattern effectiveness, directivity factor (DF), and front to back ratio (FBR) as the performance metrics. The simulation results demonstrate that the beampattern shape across different directivity patterns remains structurally consistent with quantization. However, quantization predominantly affects the SDN. Increasing the number of quantization bits increases the depth level corresponding to each null location. Furthermore, both DF and FBR were observed to be independent of the changing quantization bit depths. The results also indicate that SDN remains largely invariant across frequency, demonstrating the frequency-robust nature of null suppression in the 1D-DMA configuration. However, a notable improvement in SDN is observed with increasing quantization bits, highlighting the beneficial impact of higher quantization precision on interference rejection capability. The study also explores the effect of steering the null position from 0° to 180°. It was observed that as the null direction approaches the source look direction (0°), the depth upto which a null gets suppressed decreases, indicating reduced suppression capability. So, properly setting the null positions is important for DMA design, which plays an important role on performance. Overall, these findings provide insights about the signal quantization effects on performance measures of 1D-DMA.